\documentclass[journal, twocolumn]{IEEEtran}

\usepackage{cite}

\ifCLASSINFOpdf
  \usepackage[pdftex]{graphicx}
\else
  \usepackage[dvips]{graphicx}
\fi

\usepackage[cmex10]{amsmath}

\usepackage{algorithmic}

\usepackage{array}
\usepackage{multirow}

\ifCLASSOPTIONcompsoc
 \usepackage[caption=false,font=normalsize,labelfont=sf,textfont=sf]{subfig}
\else
 \usepackage[caption=false,font=footnotesize]{subfig}
\fi

\usepackage{stfloats}

\ifCLASSOPTIONcaptionsoff
 \usepackage[nomarkers]{endfloat}
\let\MYoriglatexcaption\caption
\renewcommand{\caption}[2][\relax]{\MYoriglatexcaption[#2]{#2}}
\fi

\usepackage{url}

\usepackage{orcidlink}

\usepackage{siunitx}
\DeclareSIUnit{\ppmm}{ppmm}

\begin{document}

\title{Multi-Platform Methane Plume Detection via Model and Domain Adaptation}

\author{Vassiliki Mancoridis\orcidlink{0000-0001-5958-7211}, Brian Bue \orcidlink{0000-0002-7856-3570}, Jake H. Lee \orcidlink{0000-0002-2838-2878}, Andrew K. Thorpe \orcidlink{0000-0001-7968-5433}, Daniel Cusworth \orcidlink{0000-0003-0158-977X}, \newline Alana Ayasse \orcidlink{0000-0001-8401-1185}, Philip G. Brodrick \orcidlink{0000-0001-9497-7661}, and Riley Duren \orcidlink{0000-0003-4723-5280} 
\thanks{\emph{(Corresponding author: Vassiliki Mancoridis.)}}
\thanks{Vassiliki Mancoridis is with the Department of Civil and Environmental Engineering, Massachusetts Institute of Technology, MA 02139 USA. This work was completed at the Jet Propulsion Laboratory, California Institute of Technology, Pasadena, CA 91109 USA (email: vm8@mit.edu)}
\thanks{Brian Bue, Jake H. Lee, Andrew K. Thorpe, and Phil G. Brodrick are with the Jet Propulsion Laboratory, California Institute of Technology, Pasadena, CA 91109 USA (email: bbue@jpl.nasa.gov; jake.h.lee@jpl.nasa.gov; andrew.k.thorpe@jpl.nasa.gov; philip.brodrick@jpl.nasa.gov)}
\thanks{Daniel Cusworth, Alana Ayasse, and Riley Duren are with Carbon Mapper, Pasadena, CA 91105 USA (email: dan@carbonmapper.org, alana@carbonmapper.org, riley@carbonmapper.org)}

}

%
%

\markboth{Preprint submitted to IEEE TRANSACTIONS ON GEOSCIENCE AND REMOTE SENSING}
{Mancoridis et al. \MakeLowercase{\textit{et al.}}: Multi-Platform Methane Plume Detection via Model and Domain Adaptation}
%



\maketitle

\begin{abstract}
Prioritizing methane for near-term climate action is crucial due to its significant impact on global warming. Previous work used columnwise matched filter products from the airborne AVIRIS-NG imaging spectrometer to detect methane plume sources; convolutional neural networks (CNNs) discerned anthropogenic methane plumes from false positive enhancements. However, as an increasing number of remote sensing platforms are used for methane plume detection, there is a growing need to address cross-platform alignment. In this work, we describe model- and data-driven machine learning approaches that leverage airborne observations to improve spaceborne methane plume detection, reconciling the distributional shifts inherent with performing the same task across platforms. We develop a spaceborne methane plume classifier using data from the EMIT imaging spectroscopy mission. We refine classifiers trained on airborne imagery from AVIRIS-NG campaigns using transfer learning, outperforming the standalone spaceborne model. Finally, we use CycleGAN, an unsupervised image-to-image translation technique, to align the data distributions between airborne and spaceborne contexts. Translating spaceborne EMIT data to the airborne AVIRIS-NG domain using CycleGAN and applying airborne classifiers directly yields the best plume detection results. This methodology is useful not only for data simulation, but also for direct data alignment. Though demonstrated on the task of methane plume detection, our work more broadly demonstrates a data-driven approach to align related products obtained from distinct remote sensing instruments.
\end{abstract}
\begin{IEEEkeywords}
Transfer learning, generative adversarial network, remote sensing, machine learning, methane plume detection.
\end{IEEEkeywords}

%
\IEEEpeerreviewmaketitle

\section{Introduction}

\begin{figure}[t]
\centering
\includegraphics[width=0.9\columnwidth]{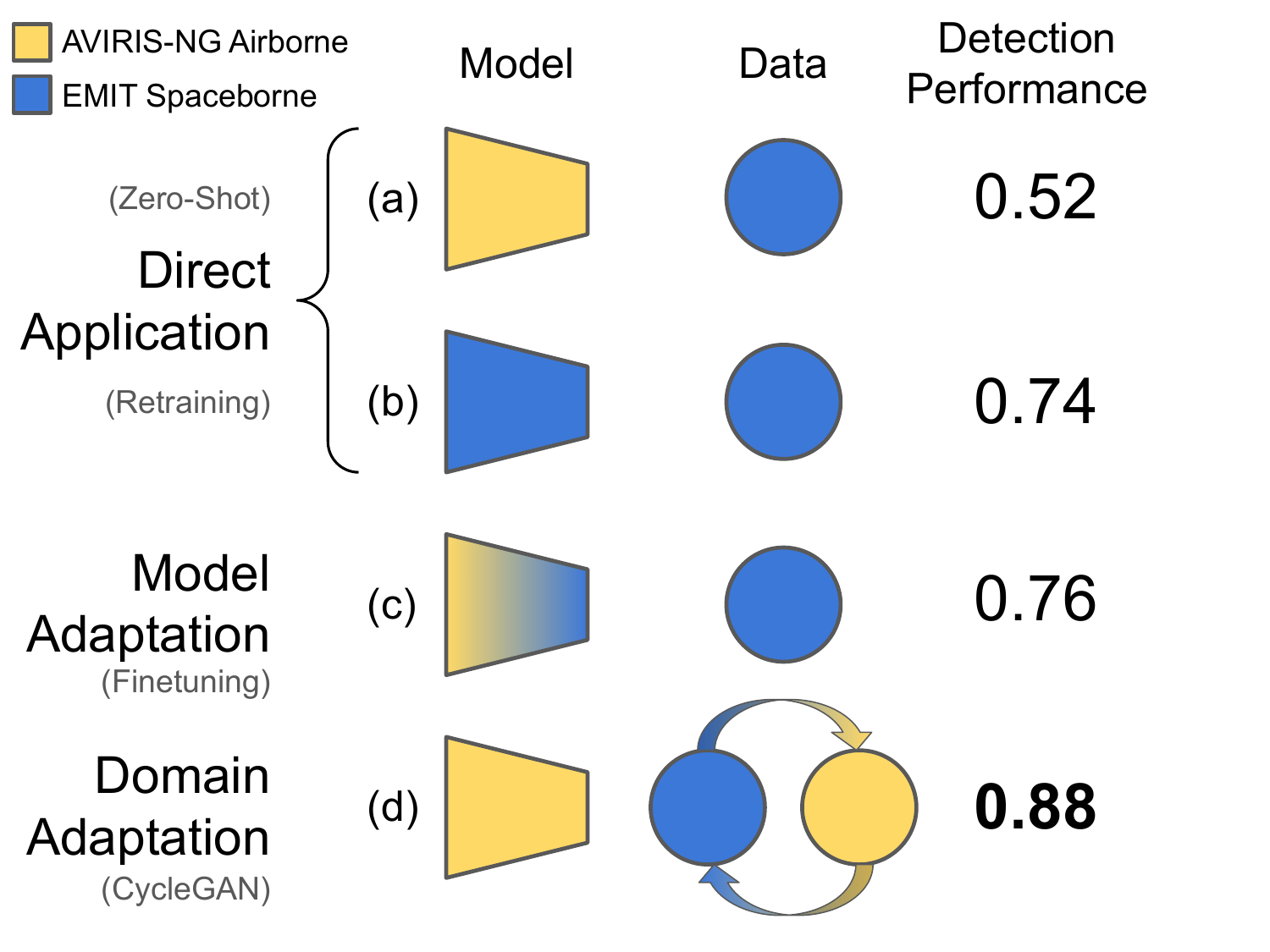}
\caption{Performance diagram summarizing approaches to developing a methane plume detection model for spaceborne EMIT data. (a) applies the airborne AVIRIS-NG model directly on the spaceborne data, and (b) applies a model only trained on the smaller amount of spaceborne data. (c) adapts the airborne model into a spaceborne model with transfer learning and fine-tuning. Finally, (d) applies the airborne model on spaceborne data adapted to the airborne domain with CycleGAN.}
\label{fig:fig1_diagram}
\end{figure}

\IEEEPARstart{M}{ethane} is a potent greenhouse gas with significant anthropogenic sources. Given its high climate forcing potential, identifying and mitigating highly disproportionate emission sources-- large plume releases-- is a priority for near-term climate action. Until recently, the prevailing strategy to measure methane emissions at global scale has been the use of area flux mappers, such as GOSAT, TROPOMI, and OCO-2 \cite{yokota2009global, hu2018toward, maksyutov2018developing}. More recently, however, imaging spectroscopy has emerged as a technique that can reliably estimate methane emissions at much finer spatial resolutions, which can be used directly for mitigation efforts. This effort began in earnest through the use of NASA Jet Propulsion Laboratory's Airborne Visible/Infrared Imaging Spectrometer \cite{green1998imaging} to map terrestrial methane sources using a columnwise matched filter \cite{thorpe2013high}, and has since expanded to multiple large-scale imaging spectroscopy campaigns \cite{cusworth2021intermittency}. The recent EMIT imaging spectroscopy mission took this process to orbit operationally \cite{thorpe2023attribution}, with over 1,400 methane super-emitters identified across six continents to date \cite{green2023emit}. As demonstrated, utilizing remote sensing products for environmental risk management works well in the context of targeted, single-instrument campaigns.

However, the issue of cross-platform methane plume detection has not yet been addressed. This is especially pertinent as the suite of orbital and airborne imaging spectrometers continues to grow. For example, the Carbon Mapper Coalition's Tanager satellite instruments precisely detect carbon dioxide, methane, and other trace gases \cite{zandbergen2023preliminary}, while MethaneSAT targets the oil and gas industries for spaceborne methane detection \cite{rohrschneider2021methanesat}. As more instruments are designed for methane detection, it is increasingly important to reconcile distributional differences between platforms. 

Prior plume detection work mostly involves training and testing on observational data from a single instrument, without introducing domain shift. For instance, Radman et al. developed a large dataset using Sentinel-2 satellite imagery and use it to train various deep learning models to estimate methane source rates \cite{radman2023s2metnet}. Lee et al. trained a convolutional neural network for plume detection and classification using multi-campaign data from the AVIRIS-NG instrument \cite{lee2022robust}. Moreover, Schuit et al. introduced a two-step machine learning approach using a convolutional neural network and a vector classifier in order to identify methane plumes from a large data source of TROPOMI observations \cite{schuit2023automated}. Finally, Bue et al. presented plume detection models with operational capabilities for AVIRIS-NG and EMIT separately \cite{bue2025towards}. These contributions harness single-instrument observations with a focus on machine learning methods for retrieval and classification tasks. 

There are few machine learning driven instances of cross-domain or cross-sensor methane plume detection. The most common approach is to train a model using synthetic Weather Research and Forecasting Model-Large Eddy Simulation (WRF-LES) generated plumes that are additively injected into plumeless background observations captured by a particular remote sensing platform. For instance, MethaNet uses WRF-LES simulated point source plumes superimposed on real AVIRIS-NG scenes to both infer wind speed and emission rates, as well as construct a prototype plume detector \cite{jongaramrungruang2022methanet}. Similarly, Joyce et. al use the WRF-LES technique to generate a large sample of simulated methane plume scenes using data from the PRISMA instrument; using these scenes as positive samples, they train a neural network to detect plumes \cite{joyce2022using}. One instance of cross-sensor methane plume detection is the work of Růžička et. al, in which the authors train a convolutional neural network using data from the AVIRIS-NG spectrometer and directly apply it to EMIT data, without adaptation \cite{ruuvzivcka2023semantic}. These approaches develop models based on training distributions fundamentally different from the target distributions. 

The common shortcoming of these techniques is that they do not account for distribution shifts effectively. A distribution shift in this context refers to the differences in the spreads of methane concentrations (ppmm values) between different remote sensing platforms due to variations in sensor resolution, altitude, and environmental conditions. Current techniques fall short in adapting to the variability in real-world data. For example, models trained on WRF-LES simulated plumes are unlikely to generalize well to real data due to oversimplification in the simulation process or incomplete treatment of background noise. The synthetic data is not ideal for capturing diffuse sources or scenes with false enhancements. In addition, approaches that use WRF-LES often do not devote significant effort to evaluating their detectors on real observations. Zero-shot approaches such as the direct application of a model trained on AVIRIS-NG to EMIT data do not address the domain drift inherent with inferencing across different remote sensing platforms. Addressing distributional shifts is essential for improving the generalizability and accuracy of methane plume detection models.

One instance of cross-sensor methane plume detection is the work of Růžička et. al, which proposes new Transformer-based architectures (HyperspectralViTs) that operate directly on raw hyperspectral radiances without relying on classical products like matched filters. Their end-to-end models achieve improved performance and on-board efficiency, with applications to both methane detection and mineral identification. Similarly to this work, the fine-tuning approach outperforms simple zero-shot generalization. In contrast, our approach addresses the challenge of distribution shift between sensors and demonstrates how domain translation methods can enhance cross-sensor generalization. Together, these works highlight complementary strategies: one aiming to build hyperspectral-native architectures for efficient on-board inference, and the other focusing on adapting existing models across sensing platforms using transfer learning techniques. Other sources that leverage transfer learning for remote sensing and Earth observation may be found in Ma et. al \cite{ma2024transfer}.

Our approach is also conceptually aligned with Mateo-García et al. \cite{ruuvzivcka2023semantic}, who use CycleGAN-based domain adaptation to improve cloud detection transfer from Landsat-8 to Proba-V. Both works address the challenge of domain shift across remote sensing instruments using unpaired image-to-image translation. However, while their study focuses on improving segmentation performance for cloud detection using spatial-spectral upscaling and a dedicated segmentation-consistency loss, our work focuses on the more fine-scale task of methane plume detection, demonstrating that domain translation alone—without paired observations—can substantially improve performance across airborne and spaceborne sensors. Other works that leverage generative adversarial networks for remote sensing may be found in a comprehensive review paper written by Jozdani et. al \cite{jozdani2022review}.

The novel contribution of this work is to introduce machine learning driven methodology that leverages the data obtained from AVIRIS-NG airborne campaigns to improve spaceborne methane plume detection from the EMIT imaging spectroscopy mission. There are two approaches that can be used to address the distribution shift issue between these platforms: adapting existing \emph{models} or adapting existing \emph{data sets}. Here, we investigate both, starting by employing transfer learning \cite{torrey2010transfer} to adapt the weights of CNNs already trained on data from one domain (airborne data) to perform well on data from a similar but related domain (spaceborne data). In contrast, we experiment with coupled generative adversarial networks capable of learning transfer functions between the two data distributions and simulating data across their distribution shift. One benefit the generative adversarial network approach is that it eliminates the need for co-registration of data products-- such as aligning locations or plume sources-- from different sensors. These domain translation techniques are especially valuable in cases with a relatively large dataset from one domain and a relatively smaller dataset from another. Our application, with nearly 2,000 labeled positive events from AVIRIS-NG and 327 labeled positive events from EMIT at the time of this work, is therefore well suited for such approaches. To the best of the authors’ knowledge, this is the first study to attempt using data-driven machine learning approaches to enhance spaceborne methane plume detection.

We find that our techniques have significantly improved performance over direct application baselines. Figure \ref{fig:fig1_diagram} illustrates our findings. A zero-shot direct application of a model trained solely on AVIRIS-NG data and tested on the EMIT data has an $F_1$ performance of 0.52. Another baseline that exclusively trains and tests a model on EMIT data has a performance of 0.74. Our model adaptation and domain adaptation approaches, which leverage both the airborne and spaceborne data, have improved performances over our baselines: 0.76 and 0.88, respectively. 

This article is organized as follows: In Section \ref{sec:data_methods}, the AVIRIS-NG and EMIT data platforms are introduced, and the model adaptation and domain adaptation approaches are stated. In Section \ref{sec:findings}, results for spaceborne dataset curation, airborne-spaceborne distribution shift quantification, direct cross-domain applications, as well as various machine learning approaches, are given. The concluding remarks are presented in Section \ref{sec:conclusion}.

\section{Data and Methods}
\label{sec:data_methods}

In this work, we derive pixel-wise greenhouse gas (GHG) concentrations from calibrated imaging spectrometer radiance observations using the Column-wise Matched Filter (CMF) approach described in Thompson et al. \cite{thompson_real_2015}. Analogously to Villeneuve et al. \cite{villeneuve1999improved}, the CMF retrieval computes pixel-wise enhancements of visible to shortwave infrared (VSWIR) absorption features relative to a lab-measured GHG transmittance spectrum versus a multivariate (channel-wise) Gaussian background distribution. Each background distribution is estimated column-wise using the observations captured by each detector element in the sensor’s focal plane array.
 
CMF products from the AVIRIS-NG instrument \cite{chapman2019spectral} were derived using the approach described in Thompson et al. \cite{thompson_real_2015}, which retrieves pixel-wise CH4 enhancements relative to the depth of the SWIR2 CH4 absorption features observed in the airborne measured radiances captured at ground sample distance (GSD) $\in [3,7]$ m\textsuperscript{2}. The CMF products of the EMIT instrument \cite{thompson2024orbit} are derived using the same method, but are computed using all VSWIR wavelengths at GSD $\approx 60$ m\textsuperscript{2} and also account for the scene-wise variation in H\textsubscript{2}O concentration and solar geometry as detailed in Thorpe et al. \cite{thorpe2023methane}.

Across both datasets, the tile size is $256\times256$ pixels. Within each tile, across both datasets, the plumes are always centered and the tiles are spatially disjoint from one another. In terms of practical implications, the difference between the spatial resolutions (GSD) of the two instruments means that a plume with a smaller footprint in the airborne dataset would be analogous to a larger, lower-resolution plume in the EMIT domain. The main challenge in aligning the two domains is to match the noise characteristics observed by the different sensors along with the methane signal. For each dataset, we partition samples into training, validation, and test sets, adhering to an allocation ratio of $80\%$, $10\%$, and $10\%$, respectively.

\subsection{Airborne dataset curation}
\label{subsec:airborne-data-curation}

For the airborne component, we utilize a large, curated dataset consisting of nearly 2,000 labeled methane plumes and 25,000 background samples, described in detail by \cite{bue2025towards}. Tiles of size $256\times256$ were sampled from 500 AVIRIS-NG images from 100 flight-lines covering regions primarily in the Southwest US. There were three airborne campaigns that comprised the data: CACH4 (California, 2018) \cite{duren2019california}, COVID (Southern California, 2020) \cite{thorpe2023methane}, and Permian (Texas Permian Basin, 2019) \cite{cusworth2021intermittency}. These tiles were split into 19,161 train tiles and 4,822 test tiles, ensuring geospatial stratification. In other words, tiles in the test set were not sampled from the same regions as tiles in the training set.

\subsection{Spaceborne dataset curation}
\label{subsec:spaceborne-data-curation}
We curate an EMIT dataset consisting of 327 labeled methane plume scenes and 8,664 background samples for training a spaceborne classifier model. We obtain the EMIT coverage map and the list of previously identified plume locations from the EMIT Open Data Portal \cite{green2023emit}. These are global samples with dates between 2022-2023, although most of the labeled plume scenes are located in the United States. This list of plume locations is geospatially matched with EMIT matched-filter-based enhancement scenes \cite{green2023emit_enh} to comprise the positive plume scenes in the EMIT dataset. Next, we construct the background set. For each EMIT scene, we first sample non-overlapping plume tiles from the boundaries of any labeled plume images. We then sample background tiles roughly covering the unlabeled (non-plume) regions of the image. Since methane plumes constitute a small fraction of each scene, we consider this sampling approach, along with curated scene rejection, to provide a sufficiently robust and representative background set. After tiling the dataset, we obtain 327 positive tiles and 9,820 negative tiles, split into 5,468 train tiles and 4,679 test tiles.

\begin{figure}[t]
\centering
\subfloat[]{\includegraphics[width=1.15in]{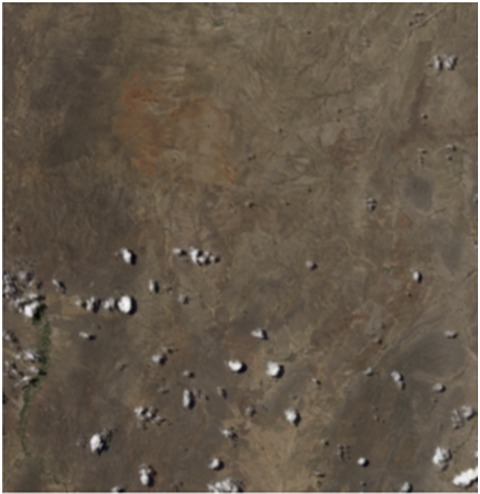}%
\label{fig_first_case_em}}
\hfil
\subfloat[]{\includegraphics[width=1.15in]{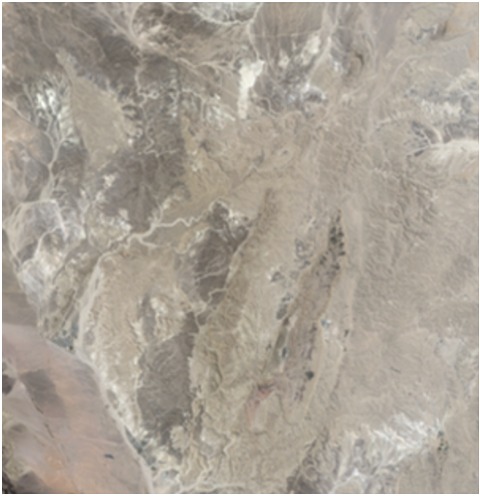}%
\label{fig_second_case_em}}
\hfil
\subfloat[]{\includegraphics[width=1.15in]{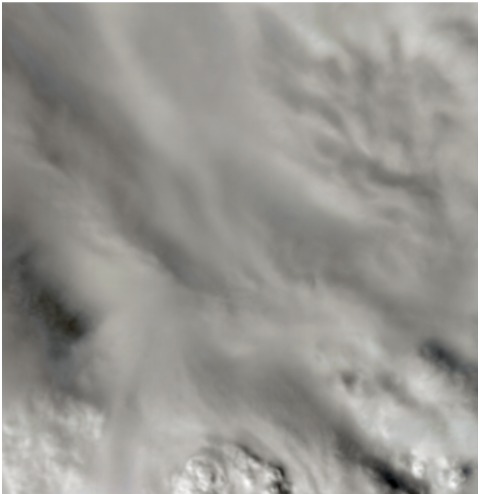}%
\label{fig_third_case_em}}
\caption{Illustration of cloud coverage in EMIT image tiles. (a) Cloud fraction of $10\%$. (b) Cloud fraction of $43\%$. (c) Cloud fraction of $100\%$.}
\label{fig_cloud_coverage}
\end{figure}

We run the operational EMIT cloud detector \cite{thompson2014rapid} on the tiles of our dataset as a means to filter out cloudy scenes that can compromise its integrity. Some example tiles from this set, with associated cloud fractions, are shown in Figure \ref{fig_cloud_coverage}. While most scenes are reliably labeled, some are less clear. For instance, in Figure \ref{fig_second_case_em}, visual inspection of the RGB image indicates that the scene contains either optically thin cirrus, or is mislabeled. In either case, methane matched filter signatures can likely be reliably obtained from within the scene. To avoid possible effects of over-reliance on the cloud detector, we define a low cutoff for labeling a scene as cloudy. If the cloud detector assigned it a cloud fraction of $20\%$ or higher, we label the scene as cloudy and reject it from our dataset. Although some cloudless scenes are lost in this process, we are able to ensure that the resulting dataset was relatively cloudless. This new, cloudless dataset has 327 positive tiles and 8,664 negative tiles.

\subsection{Direct application}
\label{train_space}

We calculate the distribution shift between the airborne and spaceborne data, focusing specifically on three datasets: the original (cloudy) EMIT dataset, the cloudless EMIT dataset, and the multi-campaign airborne dataset, which consists of data from the CACH4, COVID, and Permian campaigns.

To mitigate the effect of outliers on the results and to filter out background noise, all three datasets are clipped at their 95th percentile and at 0, discounting negative values. Since the goal of this work is to assess cross-sensor plume detection capabilities, we focus our models on pixels that are both useful for separating plumes from background or false enhancements and also informative enough to drive the CycleGAN optimization.

Generally speaking, plumes should have higher CMF-derived pixel-wise values ($v_{\text{orig}}$) than background enhancements, but the range of values most relevant to measuring methane plumes varies with respect to instrument capabilities and observing conditions. A natural approach is to filter out background noise pixels with enhancements below a minimum threshold to simplify manual plume identification, so we simply exclude values below $0$. If a pixel has a negative value, it means that it is uncorrelated with the methane enhancement modeled by the CMF, indicating that there cannot be methane there in principle. However, we still want to capture the dynamic range of the instrument, so we compute the instrument-specific scalar upper bound maximum ($\text{instr}_{\text{max}}$) from the 95th percentile of the collective pixel values observed in all scenes captured by each instrument used in our training and test sets. All scenes are then scaled into the [0,1] range to obtain a normalized value ($v_{\text{norm}}$) according to $v_{\text{norm}} = \text{CLIP}(v_{\text{orig}}, (0, \text{instr}_{\text{max}})) / {\text{instr}_{\text{max}}}$. We note that $\text{instr}_{\text{max}}$ should not be set to the maximum observed value per instrument since the normalized images would be skewed with respect to instrument specific outliers and would distort---rather than help reconcile---pixel-wise comparisons across sensors. 

Table \ref{tab:summary_stats} quantifies the differences between the three distributions after the initial clipping. The multi-campaign airborne dataset has a lower mean and maximum than either of the EMIT datasets. With a lower lower variance and range, the distribution of the EMIT cloudless dataset is lower and more condensed than the distribution of the EMIT cloudy dataset. This is because clouds are false enhancements that considerably skew the variance.

\begin{table}[h]
\caption{Summary statistics of datasets calculated after clipping to 0th and the 95th percentile.}
\label{tab:summary_stats}
\centering
\begin{tabular}{|l|r|r|r|}
\hline
Statistic & Airborne & EMIT cloudy & EMIT cloudless \\
\hline
Average (ppmm) & 127.4 & 183.6 & 158.9 \\
Maximum (ppmm) & 926.6 & 8285.0 & 1164.8 \\
\hline
95th Percentile (ppmm) & 550.7 & 808.9 & 705.8 \\
Variance (ppmm) & 38887.4 & 123267.9 & 62218.6 \\

\hline
\end{tabular}
\end{table}

For the direct application approach, we train baseline classifiers on each dataset and applied inference across distributions. Namely, we train separate classifiers on the EMIT cloudless data; on the CACH4, COVID, and Permian campaigns; on only the CACH4 and COVID campaigns; and on just the COVID campaign, which is considered a relatively small and clean dataset with low signal-to-noise ratio. Each classifier uses the GoogLeNetAA architecture with antialiasing \cite{zhang2019making}, which builds upon the original GoogLeNet work \cite{szegedy2015going}. This architecture uses parallel filters of multiple sizes and pooling within each model to capture information at multiple scales, and approximates a sparse structure using dense components that are hardware-efficient. The use of multiple filter sizes and pooling gives the model more robustness to noise and a stronger ability to recognize objects that appear at different scales between tiles. We train each GoogLeNetAA model with a learning rate of \num{1e-4} and a batch size of 16 for 100 epochs. Each GoogLeNetAA classifier had approximately 13.4 million learnable parameters (including batch normalization). We use the normalization statistics calculated via quantifying the distributions in Table \ref{tab:summary_stats}, and apply each classifier directly on data from the four distributions in order to assess cross-distributional performance. 

\subsection{Model adaptation}

In cases where data collection is limited, instead of training a randomly initialized model on relatively few examples, it is often useful to start with a model that is already trained on data from a different, but related, domain. This concept, called transfer learning, is a widely used method of model adaptation in the machine learning community \cite{weiss2016survey}. The amount of data needed for fine-tuning a model via transfer learning depends on the similarity between the source and target domains. 

Certain intuitions about data representations in convolutional neural networks can improve model adaptation results. The depth of the model and the feature space representation of each layer can influence the model adaptation process, especially when freezing and unfreezing layers. One main idea behind feature representation in neural networks is that deeper layers capture more precise features \cite{8474912}. While transfer learning, one can unfreeze the entire source domain-trained network, allowing backpropogation to flow freely through the whole model. Alternatively, certain layers can be kept frozen, particularly more shallow ones, letting others adapt to the target domain. This approach directs optimization efforts more effectively towards model areas most divergent from the target domain, enabling these deeper layers to become more specialized for the target task. However, this technique can actually worsen performance if there is co-adaptation between neighboring layers, and the freeze-unfreeze split divides these layers \cite{yosinski2014transferable}. These intuitions are crucial to fine-tuning using model adaptation techniques. 

For the model adaptation approach, we explore the effectiveness of transfer learning to fine-tune a high performing multi-campaign airborne model to the EMIT data. To maximize the efficiency of this process and avoid co-adapted neighboring layers, we investigate the effect of fine-tuning airborne models while freezing various model layers during transfer learning. 

Namely, we employ transfer learning by starting with the best performing airborne model---the three-campaign model---and refining it further on spaceborne data. Fine tuning a model with transfer learning involves unfreezing some or all of its layers during the additional learning process. The GoogLeNetAA model architecture used for the classifiers consists of a series of inception blocks and some auxiliary branches for gradient preservation \cite{szegedy2015going}. We run experiments in which we unfreeze different inception blocks (i.e. layers) of the model before further training on the spaceborne data. Our guiding intuition was that the inception blocks at the beginning of the network should capture the general essence of a plume since they represent broad features. Plume characteristics are expected to be consistent across airborne and spaceborne distributions. With a limited number of epochs for fine-tuning, we suspect that the model may optimize better to the spaceborne distribution if it focuses its additional training only on deeper network layers, which represent more granular features. In addition, training on only the deeper network layers may prevent this new, fine-tuned model from overfitting to the additional EMIT cloudless dataset on which it trained. We also seek the optimal number of layers to unfreeze when adapting an airborne model to spaceborne data. As such, for each configuration of freezing and unfreezing subsequent layers, we run additional training using the cloudless EMIT data for 20 epochs. 

\subsection{Data adaptation}

\begin{figure*}[t]
\centering
\includegraphics[width=1\linewidth]{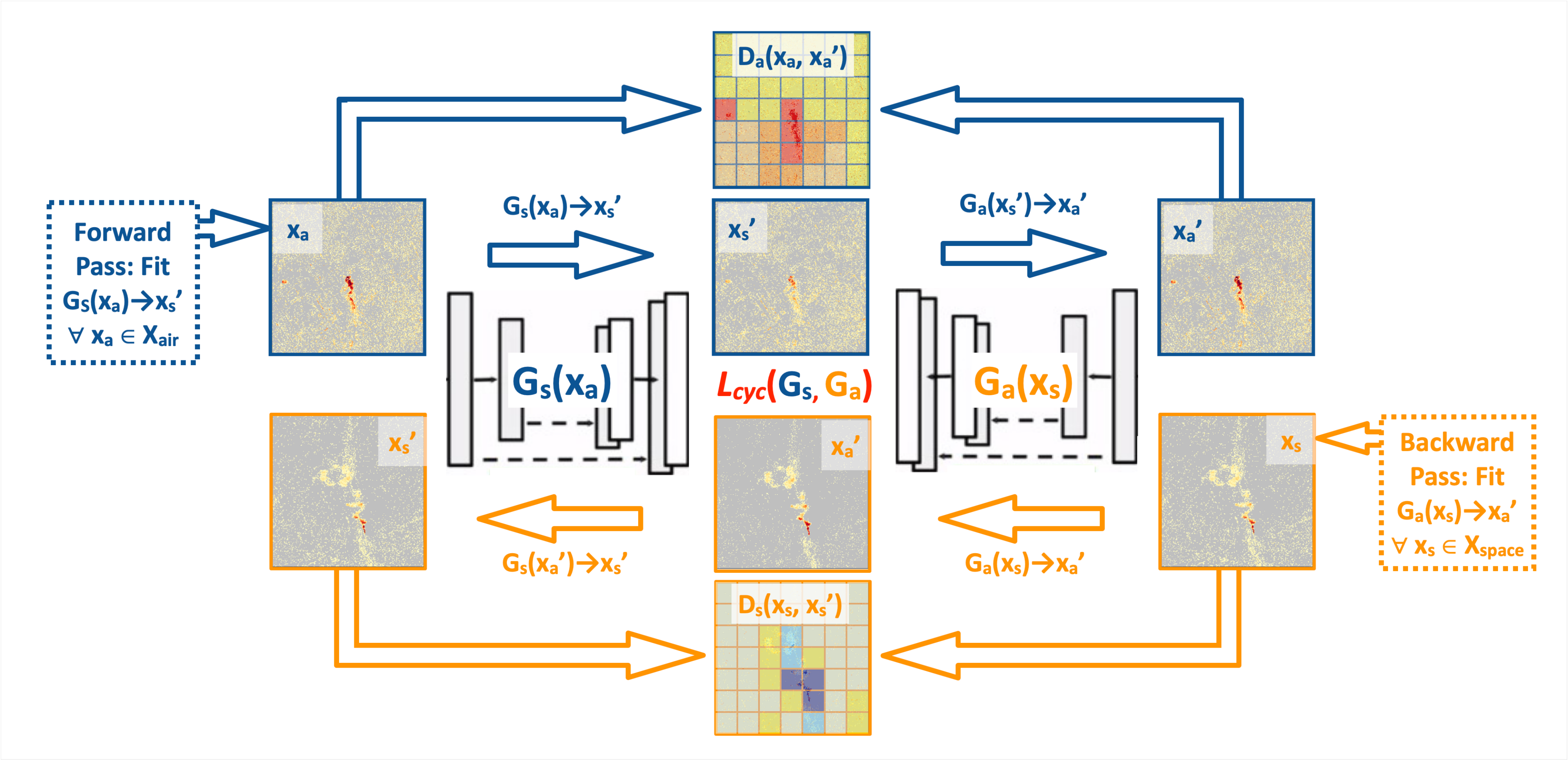}
\DeclareGraphicsExtensions.
\caption{Flowchart illustrating CycleGAN optimization to learn a bidirectional mapping between airborne (AVIRIS-NG) samples $x_{a} \in X_{\text{air}}$ and spaceborne (EMIT) samples $x_{s} \in X_{\text{space}}$. CycleGAN learns this mapping by jointly optimizing a pair of conditional GANs ($G_{a}, G_{s}$) that each map samples from one domain to the other, while discriminators $D_{a}(x_{a},x'_{a})$ and $D_{s}(x_{s},x'_{s})$ are optimized to distinguish real samples from samples produced by the conditional GANs. The cyclic consistency loss $L_{\text{cyc}}$ guides the optimization by penalizing each generator if it produces cyclic disparities (e.g., $G_{a}$ produces outputs that $G_{s}$ cannot map back to the $X_{\text{air}}$ domain, and vice versa). Blue arrows and annotations indicate the "forward pass" optimization which learns the $X_{\text{air}} \rightarrow X_{\text{space}}$ mapping obtained via $G_{a}$. Orange arrows and annotations indicate the "backward pass" optimization which learns the $X_{\text{space}} \rightarrow X_{\text{air}}$ mapping obtained via $G_{s}$.}

\label{cyclegan}
\end{figure*}

Generative adversarial networks (GANs) \cite{goodfellow2020generative} employ an adversarial optimization approach which iteratively refines two models, a generator and a discriminator, that jointly align training samples with generated observations \cite{creswell2018generative}. These models can be used for simulating novel data and for characterizing distributional shifts.

The adversarial optimization approach utilized in this work is the CycleGAN model \cite{zhu2017unpaired}. The Pix2Pix model \cite{isola2017image}---a component of CycleGAN---compares real and generated images against one another, analyzing images locally in patches in order to characterize differences and correct itself accordingly. As it attempts to mitigate differences between the generated and real images throughout training, Pix2Pix learns an increasingly accurate transfer function between the source and target domains. CycleGAN \cite{zhu2017unpaired} is a coupled GAN consisting of two Pix2Pix models, each undergoing adversarial training. One generator-discriminator pair focuses on the distribution shift from $X$ to $Y$, while the other focuses on the transformation from $Y$ to $X$. There is also a cycle-consistency loss term dedicated to minimizing disparities between cyclic transformations (e.g. $X \rightarrow Y \rightarrow X$). This configuration allows CycleGAN to learn bidirectional transfer functions using {\em unpaired} (i.e., not coregistered) images to bridge the gap between the source and target domains.

Figure \ref{cyclegan} depicts the CycleGAN optimization process. In the forward pass, we train conditional GAN $G_s(x_a)$ to map samples from the $x_a \in X_\text{air}$ domain to the $X_\text{space}$ domain by competing with a patchGAN discriminator $Da(x_a, x_a')$ optimized to distinguish real samples $x_a$ from their generated counterparts $x_a' = G_a(G_s(x_a))$. We use the same procedure in the backward pass to train conditional GAN $G_a(x_s)$ and patchGAN discriminator $Ds(x_s, x_s')$ to map samples from the $X_{air}$ domain to the $X_\text{space}$ domain and to separate real samples $x_s \in X_\text{space}$ from their corresponding generated counterparts $x_s' = G_s(G_a(x_s))$, respectively. Both $G_a$ and $G_s$ are optimized collectively via the cycle consistency loss $L_{cyc}(G_s, G_a)$ which drives each GAN to generate invertible predictions in the forward (i.e., $x_a \sim G_a(G_s(x_a))$) and backward (i.e., $x_s \sim G_s(G_a(x_s))$) passes. 

For the data adaptation approach, we train a CycleGAN model that has a means to map observations in the target domain (spaceborne) back to the source domain (airborne), and vice versa, thereby allowing the direct application of a source-trained classifier on target examples. The CycleGAN models are trained to translate data of two sensors with different spatial resolutions, which implies that in one direction, the model is also effectively learning superresolution, and in the other, sensor-specific down-scaling. This is beneficial for the system because it enables the model to capture and align both fine-grained structural details from high-resolution imagery and broader contextual patterns from coarser observations, thereby facilitating more robust cross-domain generalization and allowing pretrained classifiers to operate effectively across heterogeneous sensor inputs. There is no need to consider scaled data separately.

Our CycleGAN models each have approximately 28.3 million parameters; the two generators each have about 11.4 million parameters, and the two discriminators each have about 2.8 million parameters.

Our first objective is to train baseline CycleGAN models using subsets of the EMIT cloudless dataset as well as the three-campaign (CACH4, COVID, and Permian) AVIRIS-NG airborne dataset. We sample these subsets such that they contain an equal number of plume and non-plume tiles. That is, for every campaign, we subsample all positive tiles (containing plumes). Then, we randomly subsample the same number of negative tiles (no known plumes). The driving motivation behind this sampling approach is that it greatly reduces the computational cost of training the CycleGAN model. In addition, this method simplifies the CycleGAN's learning process; it allows the adversarial model to recognize plume tiles not as a set of outliers, but as a critical subsection of the distribution. Plume-to-plume cross-distributional translations are imperative because plume tiles are much more rare than non-plume tiles.

We run extensive experimentation across CycleGAN objectives, loss functions, and learning rates, both analyzing the loss trajectories and generated tiles, in order to identify a suitable and convergent CycleGAN. For the experimentation process, we train on three different objectives: the vanilla objective \cite{goodfellow2020generative}, least-squares GAN (LSGAN) \cite{mao2017least}, and the Wasserstein GAN with Gradient Penalty (WGAN-GP) \cite{gulrajani2017improved}. For each architecture, we train with a learning rate of \num{2e-3}, \num{2e-4}, and \num{2e-5}. For some objectives, we even go further and train with a learning rate of \num{2e-6}. Moreover, for the vanilla objective function, we experiment with a loss type of $L_1$ loss versus MSE loss. We train these baseline models on a small sub-sample of the airborne and spaceborne data.

Next, we devote our effort towards training larger CycleGANs that would be more useful for our data simulation tasks. Using the most effective training hyperparameters that we identified on the sub-sampled dataset, we train six larger CycleGANs on both \emph{balanced} and \emph{unbalanced} airborne datasets. Balanced datasets contain all of the positive plume tiles from each campaign and an \emph{equivalent} number of negative tiles. Unbalanced datasets contained all training tiles from each campaign. 

Another critical metric that we track throughout the CycleGANs training processes is classification $F_1$ score. After every epoch of training, we pass the airborne test set $x_a \in X_\text{air}$ through the air $\rightarrow$ space generator $G_s(x_a)$ to generate simulated spaceborne products. Then, we run these products through the EMIT baseline classifier that we trained in Section \ref{train_space}. Moreover, after every epoch of training, we also pass the spaceborne test set $x_s \in X_\text{space}$ through the space $\rightarrow$ air generator $G_a(x_s)$ to generate simulated airborne products. We run these products through the CACH4+COVID+Permian airborne model from Section \ref{direct_application_section}. For the balanced CycleGAN on the CACH4+COVID+Permian and EMIT datasets, the airborne test set contained 432 total tiles with 216 positive cases, each at \qtyrange{3}{7}{\meter} resolution; the EMIT test set contained 238 total tiles with 119 positive cases, each at \SI{60}{\meter} resolution. Passing these test sets through the respective generators alters their spatial resolutions, but do not change their positive versus negative labels. Theoretically, as the CycleGAN trains, the generators are able to learn a more accurate representation of each distribution. Therefore, we would expect the classification score for each classifier to rise. At epoch 0, no domain translation is applied—data are not passed through the CycleGAN, and models are evaluated via direct application to the target domain. Accordingly, the $F_1$ scores of the EMIT model on unaugmented airborne data, and the airborne model on unaugmented EMIT data, match the direct application values reported in Table \ref{cross_domain_baselines_table}. Subsequent epochs reflect performance on CycleGAN-translated data, capturing the effects of domain adaptation.

\section{Findings}
\label{sec:findings}

\subsection{Direct application}
\label{direct_application_section}

\begin{table*}[b]
\renewcommand{\arraystretch}{1.3}
\caption{Cross-domain Baseline $F_1$ Scores}
\label{cross_domain_baselines_table}
\centering
\begin{tabular}{|c|c|ccc|}
\hline
\multirow{ 3}{*}{} & EMIT Data & \multicolumn{3}{c|}{AVIRIS-NG Data} \\
\cline{2-5}
& Cloudless & CACh4+COVID+Permian & CACh4+COVID & COVID \\
\hline
Train Dataset Tiles & 5,469 & 19,161 & 7,959 & 4,234 \\
Test Dataset Tiles & 4,680 & 4,822 & 2,219 & 1,230 \\
$256\times256$ px. tiles, GSD & \SI{60}{\meter} & \qtyrange{3}{7}{\meter} & \qtyrange{3}{7}{\meter} & \qtyrange{3}{7}{\meter} \\
\hline
EMIT Cloudless Model & \textbf{0.74} & 0.60 & 0.59 & 0.57 \\
CACh4+COVID+Permian Model & 0.52 & \textbf{0.76} & 0.81 & 0.76 \\
CACh4+COVID Model & 0.57 & 0.71 & \textbf{0.82} & 0.85 \\
COVID Model & 0.46 & 0.64 & 0.78 & \textbf{0.92} \\
\hline
\end{tabular}
\end{table*}

The direct application results are depicted in Table \ref{cross_domain_baselines_table}. There are a few predominant findings. First, the diagonal of the table contains most of the high values. This makes intuitive sense because we would expect the models to do best when tested on similar data to that on which they were trained. The airborne models generally had weak performance on EMIT cloudless data, especially the model solely trained on the COVID campaign. Likewise, the EMIT model had trouble adapting to the airborne distribution. Next, the EMIT model was slightly better at generalizing to the airborne data than vice versa. This may be because spaceborne images are captured from a greater distance from Earth, rendering them less granular. Consequently, CNNs trained on these spaceborne images might be less prone to overfitting, enhancing their generalizability to unseen data. These domain adaptation results suggest that there is significant room for improvement via either data or model adaptation.

\subsection{Model adaptation}

\begin{figure*}[t]
\centering
\includegraphics{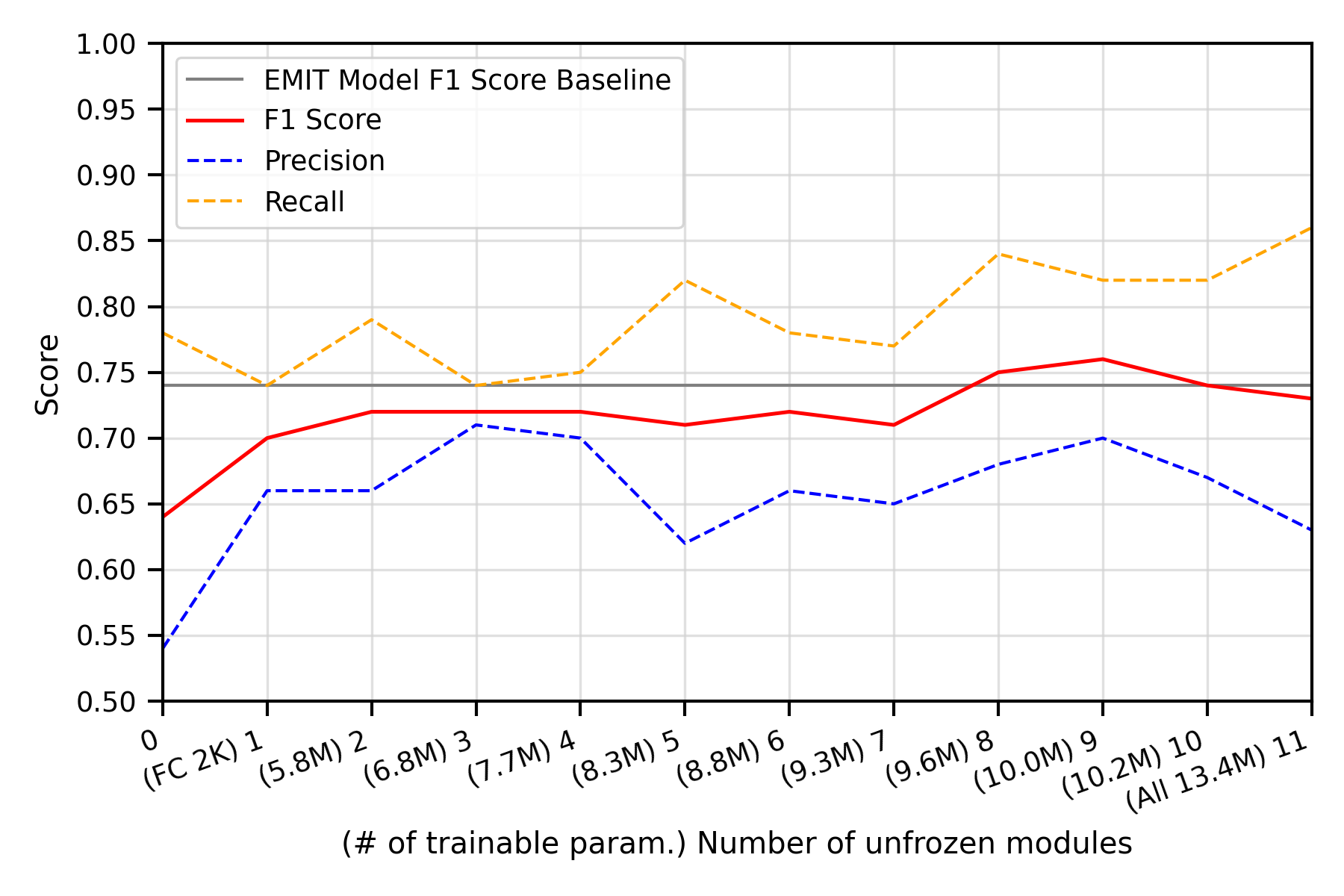}
\caption{Results of finetuning the AVIRIS-NG CACh4+COVID+Permian model on EMIT data for 20 epochs with an increasing number of unfrozen layers of the architecture during training. Numbers in parentheses indicate the number of trainable parameters; ``(FC 2K)'' indicates that only the fully-connected layer was unfrozen, only retraining 2,000 parameters.}
\label{fig:fine-tuning-results}
\end{figure*}

The model adaption results are shown in Figure \ref{fig:fine-tuning-results}. We found that unfreezing more layers increases the performance, from 0.65 with no fine-tuning to over 0.71 with any amount of fine-tuning. Moreover, as more layers were unfrozen, the model generally performed better on the spaceborne data. The model performed most optimally at nine unfrozen layers, which leaves three frozen layers at the top of the network. In this case, the airborne model refined on spaceborne data outperformed the model trained on spaceborne data alone (indicated by the gray line in Figure \ref{fig:fine-tuning-results}), though only marginally. Specifically, the precision, recall, and $F_1$ score were 0.70, 0.82, and 0.76, respectively. This is an interesting result because it suggests that the airborne data may be used to somewhat enhance spaceborne plume detection.

\begin{figure*}[t]
\centering
\includegraphics{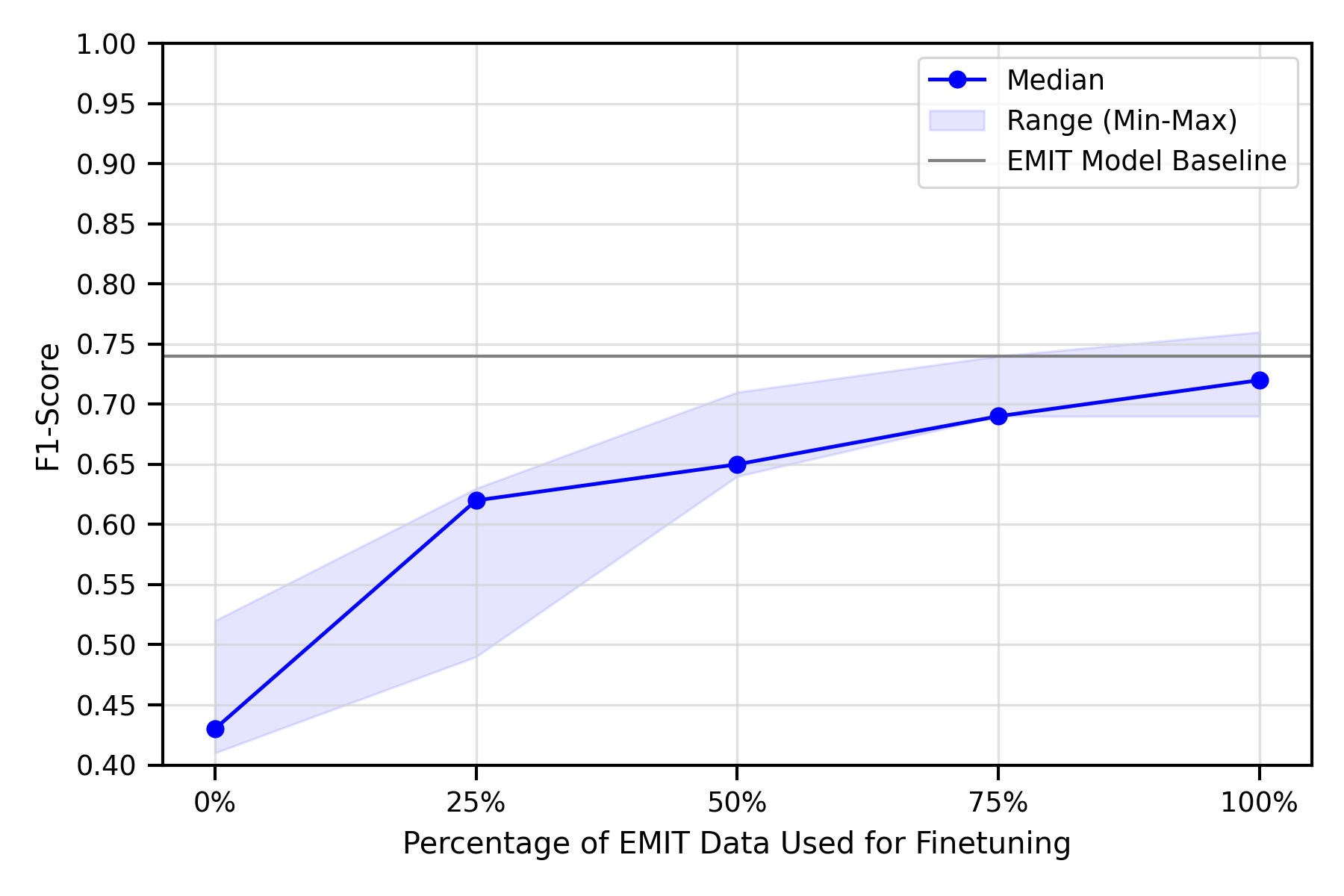}
\caption{Performance of transfer learning the CACH4-COVID-Permian AVIRIS-NG model with nine unfrozen layers on subsets of the EMIT cloudless training data. The grey line is the performance of the EMIT model on the EMIT dataset. The 0\% point is the same as a direct zero-shot application.}
\label{fig:how_low}
\end{figure*}

A natural follow-up probes how much spaceborne data is truly needed for effective additional refinement. This is critically important not just because training data is costly and challenging to obtain, but also because it takes time to acquire these data after a new sensor enters orbit; decreasing the number of samples required for fine-tune could translate into a faster transition into operations for new a new sensor. For this experiment, we kept the number of unfrozen layers fixed at the optimal number of nine, as previously determined. We re-ran the additional training, this time on varying percentages of the EMIT training data. We ran each experiment three times, across three train-test splits of the cloudless EMIT dataset, in order to eliminate sampling biases. Figure \ref{fig:how_low} depicts the results of this experiment. From these results, we see that the value of additional on-orbit training data never saturates; however, the most substantive gains are obtained with a modest 25\% of the EMIT dataset. This highlights the strong value of fine-tuning on even a relatively small dataset.

\begin{figure*}[t!]
\centering
\includegraphics[width=0.9\linewidth]{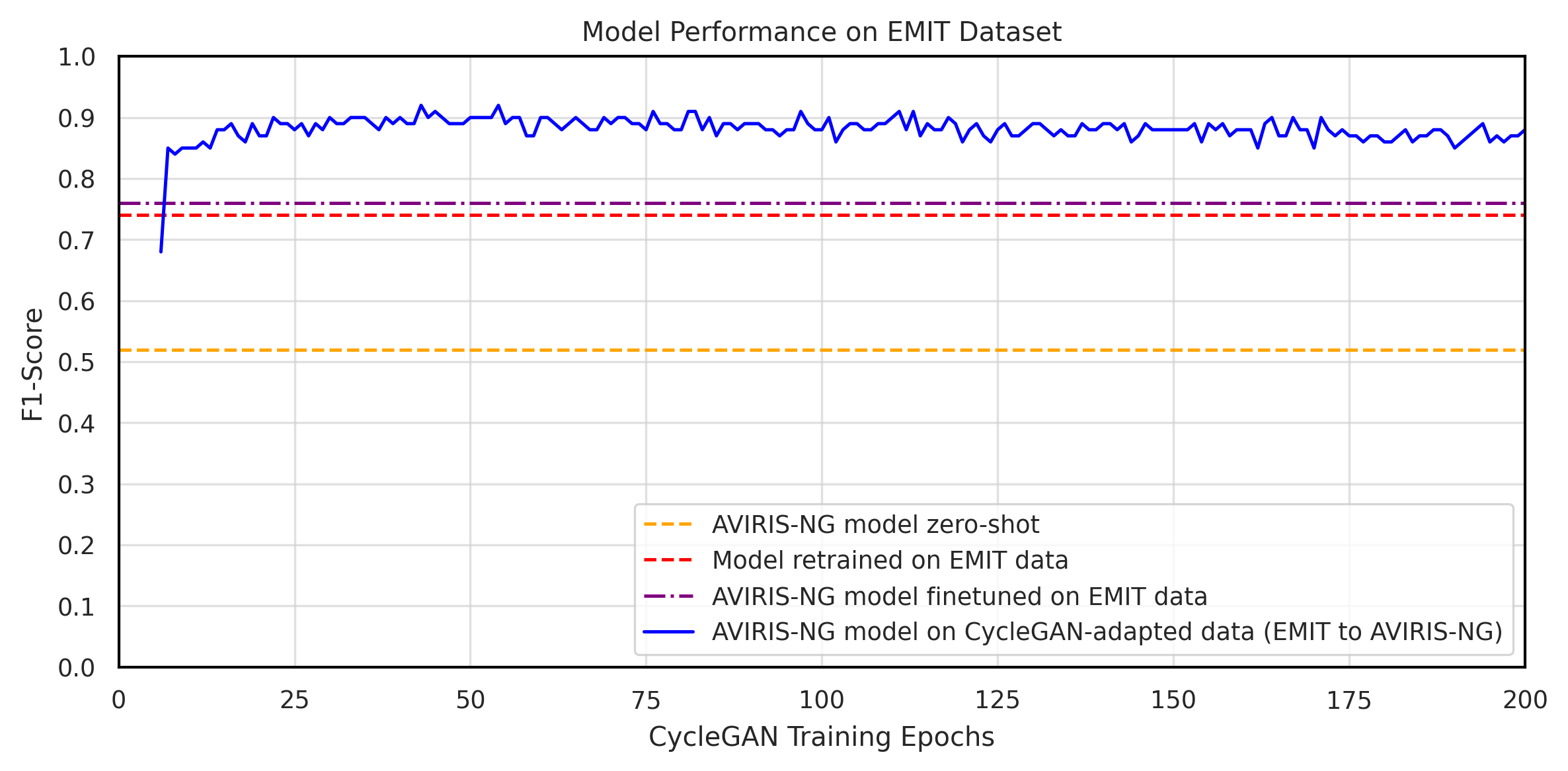}
\DeclareGraphicsExtensions.
\caption{Ability of CycleGAN to translate EMIT data to the airborne distribution as CycleGAN trains. EMIT data is passed through CycleGAN and classified using the CACH4+COVID+Permian model (blue line). Results are compared against model adaptation (purple line) and direct application (orange and red lines) baselines.}
\label{classifier_metrics_BtoA}
\end{figure*}

\subsection{Data adaptation}

Through hyperparameter tuning on a subsample of the airborne and spaceborne data, we found the most accurate model to be the CycleGAN trained on the vanilla objective with a learning rate of \num{2e-5}. As such, we trained six larger CycleGANs using learning rates between [\num{2e-5}, \num{2e-6}] on two datasets (CACH4+COVID, CACH4+COVID+Permian). The highest performing model---based on loss curves, generated images, and classifier performance---was the balanced CycleGAN that was trained on the CACH4+COVID+Permian campaigns with a learning rate of \num{2e-6}. This dataset contained 2,968 train and 432 test airborne tiles; it also had 416 train and 238 test spaceborne tiles. Exactly half of each train or test set contained positive tiles (with plumes), and half contained negative tiles (without plumes).

\begin{figure*}[t!]
\centering
\includegraphics[width=0.9\linewidth]{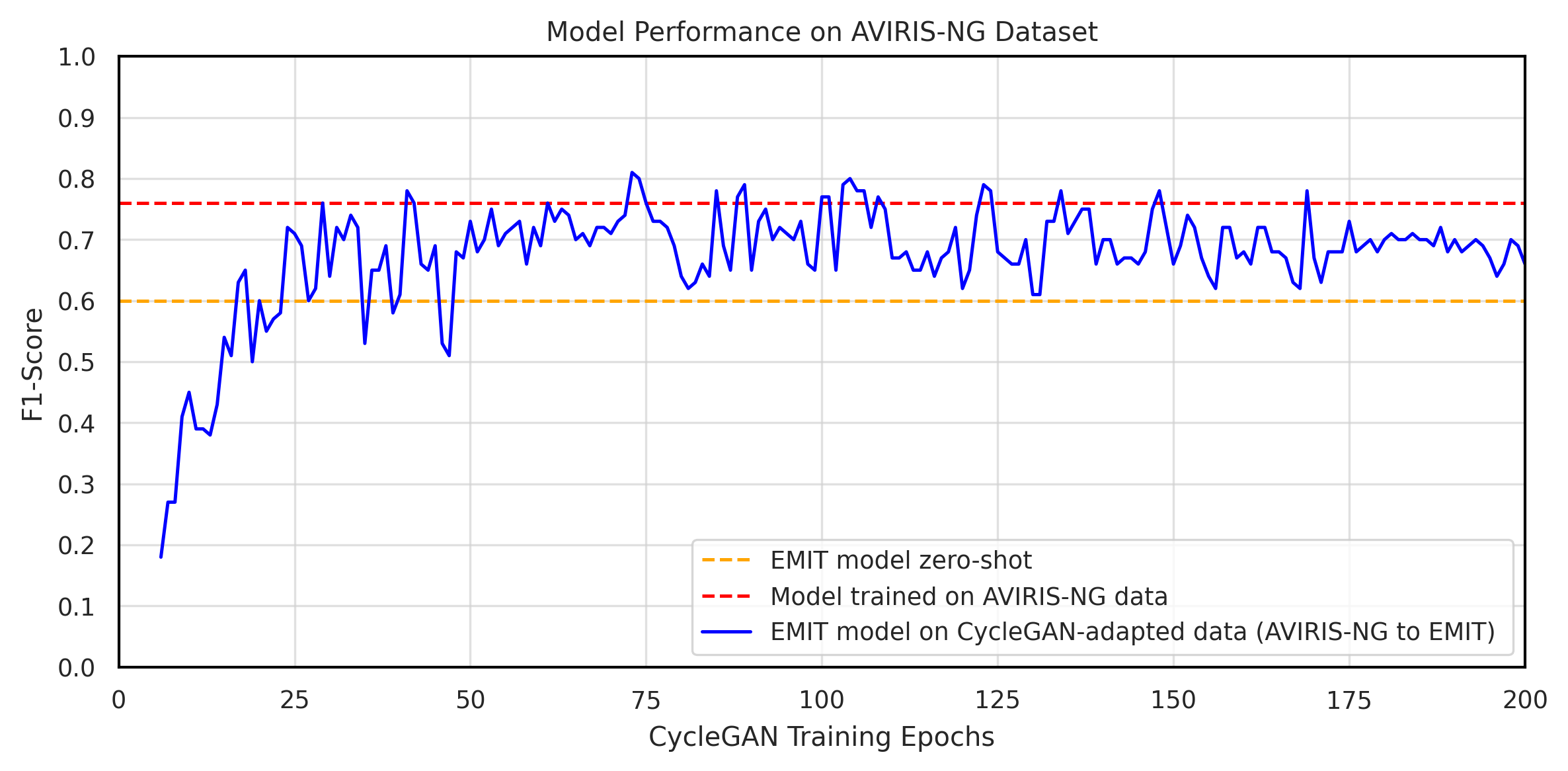}
\DeclareGraphicsExtensions.
\caption{Ability of CycleGAN to translate airborne data to the EMIT distribution as CycleGAN trains. Airborne data is passed through CycleGAN and classified using the EMIT model (blue line). Results are compared against the direct application of the EMIT model on airborne data (orange line) and of the airborne model on airborne data (red line).}
\label{classifier_metrics_AtoB}
\end{figure*}

The classification $F_1$ metrics tracked across the training of the CycleGAN model demonstrate the superiority of the data alignment approach for this classification task. The results are depicted in Figure \ref{classifier_metrics_BtoA} and Figure \ref{classifier_metrics_AtoB}. From these analyses, we conclude that the baseline classifiers perform exceedingly well on the CycleGAN-augmented products. For instance, in Figure \ref{classifier_metrics_AtoB}, notice that the $F_1$ score for the simulated EMIT products rose to about 0.7. This is about just as good as the EMIT cloudless model's performance on the held-out EMIT cloudless data, which was 0.74, as seen in Table \ref{cross_domain_baselines_table}. What is even more notable is the $F_1$ score for the simulated airborne products in Figure \ref{classifier_metrics_BtoA}. For the negative class (scenes without plumes), the $F_1$ score was 0.90 (0.83 precision, 0.99 recall); for the positive class (scenes with plumes), it was 0.88 (0.99 precision, 0.79 recall). The positive class is more difficult to classify than the negative class due to a large variation between methane plumes and the presence of false enhancements in the dataset. Yet, we see that this approach outperforms the airborne model's direct inference on the EMIT cloudless dataset, which had a score of 0.52, as seen in Table \ref{cross_domain_baselines_table}. Moreover, this result also outperformed the best transfer learned model's performance on the EMIT data, which was 0.76 (0.70 precision, 0.82 recall). This demonstrates that, out of all approaches, the spaceborne data is best classified after being transformed into the airborne distribution and then passed through our robust, multi-campaign airborne classifier.

While it is common for performance to initially dip in domain translation tasks as the model begins to reconcile differences between domains, we observe a sharp improvement after just one epoch (e.g., from $<0.7$ to $\sim 0.85$ in Fig. \ref{classifier_metrics_BtoA}). This suggests that even limited exposure to translated samples can help the model better align domain-specific features, especially when initialized with a strong pretrained backbone. Additionally, the CycleGAN may be learning coarse structural mappings early on, which are sufficient to improve task performance before finer details are refined in later epochs.

\begin{figure*}[t!]
\centering
\includegraphics[width=1\linewidth]{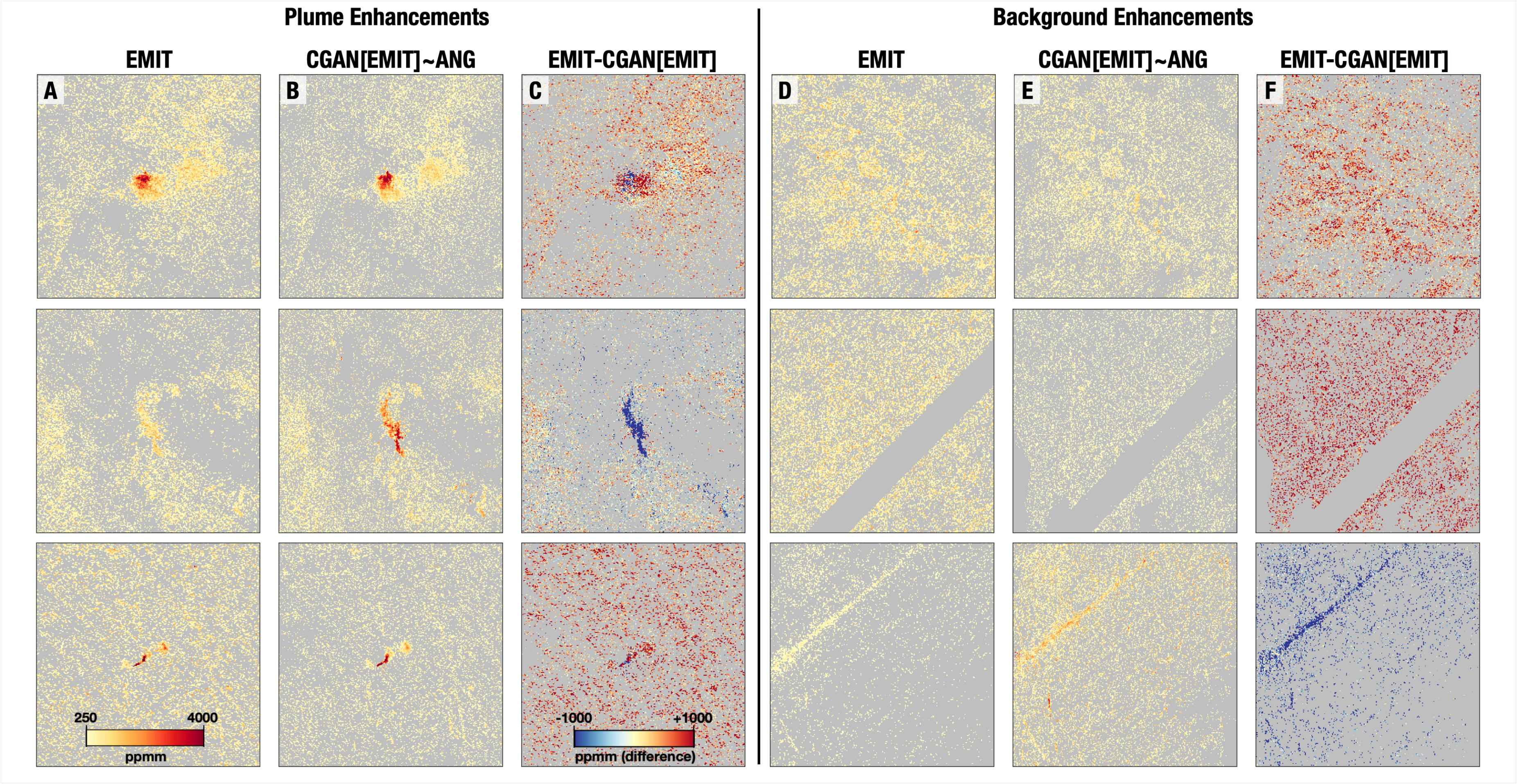}
\DeclareGraphicsExtensions.
\caption{Example EMIT plume (column A) and background (column D) tiles, their corresponding CycleGAN outputs (${CGAN[\text{EMIT}]} \rightarrow \text{ANG}'$) mapping the EMIT samples to the ANG distribution (columns B and E, respectively) along with their pixelwise differences ($\text{EMIT}$ - ${CGAN[\text{EMIT}]}$) in the \SI{\pm1000}{\ppmm} range (columns C and F, respectively). Red difference pixels indicate $\text{EMIT} > \text{ANG}'$, blue difference pixels indicate $\text{EMIT} < \text{ANG}'$, and yellow difference pixels indicate $\text{EMIT} \sim \text{ANG}'$. NODATA pixels and negative CMF enhancements in either $\text{EMIT}$ or $\text{ANG}'$ are shown in gray and are excluded from the difference maps in columns (C) and (F).}
\label{cgan_plume_bg}
\end{figure*}

Figure \ref{cgan_plume_bg} depicts the effects of passing airborne and spaceborne images through the CycleGAN models. It shows three example $x_\text{EMIT}$ plume and background (columns A and D, respectively) tiles, their generated $x_\text{ANG}'$ = $CGAN_\text{ANG}$($x_\text{EMIT}$) counterparts (columns B and E) and the pixelwise differences between each $x_\text{EMIT}$ sample and its matching $x_\text{ANG}'$ (columns C and F). The $x_\text{ANG}'$ outputs have a wider dynamic range---difference between maximum-minimum pixel values---than their $x_\text{EMIT}$ counterparts. This is evidenced by the increased $x_\text{ANG}'$ values of plume pixels (mostly blue pixels in the difference maps) and corresponding reduced values of background pixels (mostly red pixels in difference maps). This behavior is noteworthy, since as it indicates that the $CGAN_\text{ANG}$ model approximates the dynamic range and spatial variance of the fine resolution (\qtyrange{3}{5}{\meter} ground sample distance, or GSD) ANG observations from coarse resolution (\SI{60}{\meter}) EMIT observations. The dynamic range of the EMIT CMFs tends to be much narrower than the dynamic range of fine-resolution airborne CMFs captured at \qtyrange{3}{5}{\meter} GSD. This is unsurprising since methane plumes are local maxima in the CMF retrievals, so if both ANG and EMIT simultaneously observed the same plume we would expect $\max(x_\text{ANG}) > \max(x_\text{EMIT})$ due to the impact of sub-pixel mixing at \SI{60}{\meter} GSD smoothing those local maxima. This qualitative analysis enhances our understanding of how the CycleGAN model effectively simulates sensor observational characteristics during the translation of products between the airborne and spaceborne distributions.

The scale of CMF background enhancements generally follow the same trend---i.e., $\min(x_\text{ANG}) < \min(x_\text{EMIT})$ when both $x_\text{ANG}$ and $x_\text{EMIT}$ observe similar plume-free areas---except in cases where false enhancements are present. This is the case in the third example background enhancement depicted of Figure \ref{cgan_plume_bg} (bottom row). In this example, a high albedo surface feature---an asphalt road---produces a linear false enhancement. Rather than the pixel values of this false enhancement being suppressed along with the other background tiles, which is the ideal scenario for robust plume detection, the values are enhanced. This is illustrated by the blue pixels in column F, which contrast with the red pixels in the other background enhancement examples. The unexpected behavior of the $CGAN_\text{ANG}$ model on this example may be a result of the fact that linear false enhancements do occur occasionally in the EMIT retrievals, but are much more common in the ANG retrievals---which use a less sophisticated retrieval approach in comparison to the EMIT retrieval algorithm. While the CycleGAN model learns to translate products between distributions, it does not simultaneously learn to suppress false enhancements. 

\section{Conclusion}
\label{sec:conclusion}
In this work, we trained and validated an EMIT plume detector using curated background samples. We also trained plume detectors from multi-campaign airborne datasets observed by the AVIRIS-NG instrument. We evaluated zero-shot, direct application of airborne and spaceborne models across distributional shifts. Airborne models applied to EMIT data produced reasonable results. Model adaptation yielded acceptable performance, with the airborne model refined with spaceborne data marginally outperforming the standalone spaceborne model. The most compelling outcome was the use of CycleGAN to align airborne and spaceborne products. Using CycleGAN, we mapped the holdout spaceborne set into simulated airborne data. Passing this data through our strongest airborne classifier yielded the highest predictive outcome, at an $F_1$ score of 0.88 for the positive EMIT plume tiles.

At the time of this study, we lacked sufficient EMIT-only data to evaluate whether training a model from scratch on that modality would outperform domain translation approaches. As such, we prioritized rapid adaptation using AVIRIS-NG data to enable early deployment. As more EMIT data becomes available, retraining solely on EMIT or jointly on both modalities may improve performance. Even with more data labeled in the EMIT platform in the future, it is still valuable to include airborne data projected onto the spaceborne domain as a form of training data augmentation.

For the training of our CycleGAN, we undertook an unsupervised approach and did not pass in any plume labels. A caveat to our method is that we balanced the datasets such that there were an equal number of positive and negative tiles in order to simplify the learning process for the already difficult to optimize CycleGAN, so the training was not fully unsupervised.

This CycleGAN approach may offer another avenue to generate realistic, simulated plumes. Methods such as WRF-LES, widely used in the remote sensing community, are useful for simulation purposes, but are not entirely realistic \cite{varon2018quantifying, rybchuk2021statistical, bhardwaj2022evaluating}. In contrast, CycleGAN integrates plumes with background features and false enhancements, also accounting for the dynamic range characteristic to the remote sensing platform. 

In summary, our work employed deep learning to align both models and datasets in the context of the airborne-spaceborne distribution shift for methane plume detection. Out of the direct application, model adaptation, and data adaptation approaches, we had the most success with data adaptation using coupled generative adversarial networks. More broadly, this work illustrates promise in integrating such deep learning techniques within the remote sensing community, which is often challenged by distributional shifts while dealing with data collected from different sensors.

\section*{Acknowledgment}

Vassiliki Mancoridis was supported by the Summer Undergraduate Research Fellowship at the Jet Propulsion Laboratory (SURF@JPL) program sponsored by the California Institute of Technology. The research was carried out at the Jet Propulsion Laboratory, California Institute of Technology, under a contract with the National Aeronautics and Space Administration (80NM0018D0004).

The Carbon Mapper team acknowledges the support of their sponsors including the High Tide Foundation and other philanthropic donors. A portion of this work was funded by both the US Greenhouse Gas Center and the Earth Surface Mineral Dust Source Investigation (EMIT), a NASA Earth Ventures-Instrument (EVI-4) Mission. US Government Support Acknowledged.

\ifCLASSOPTIONcaptionsoff
  \newpage
\fi




%



\bibliographystyle{IEEEtran}
\bibliography{bibtex/bib/IEEEmain}

%









\end{document}